\documentclass[%
 reprint,
superscriptaddress,
amsmath,amssymb,aps,
]{revtex4-2}

\usepackage{placeins}
\usepackage{graphicx}
\usepackage{dcolumn}
\usepackage{bm}
\usepackage{siunitx}
\usepackage{braket}
\usepackage{verbatim}

\DeclareSIQualifier\milliwattref{m}
\DeclareSIUnit\dbm{\decibel\milliwattref}

\begin{document}

\preprint{APS/123-QED}

\title{Comparing optical--microwave conversion and all‑microwave control schemes for a transmon qubit}

\author{Volodymyr Monarkha}
\author{Massimo Borrelli}
\affiliation{Bluefors Oy, Arinatie 10, 00370 Helsinki, Finland}
\author{Reza Hajitashakkori Kenari}
\author{Mohammad Kobba}
\author{Eugenio Cataldo}
\author{Beer de Zoeten}
\author{Mahnaz Zarrinfar}
\author{Kamal Pandey}
\author{Abhinand Pusuluri}
\author{Filippo D. Michelacci}
\author{Eliot Jouan}
\author{Bennett Sprague}
\author{Simon Groeblacher}
\author{Thierry C. van Thiel}
\author{Robert Stockill}\email{rob@qphox.eu}
\affiliation{QphoX B.V., Elektronicaweg 10, 2628XG, Delft, The Netherlands}
\author{Russell E. Lake}\email{russell.lake@bluefors.com}
\affiliation{Bluefors Oy, Arinatie 10, 00370 Helsinki, Finland}

\date{\today}

\begin{abstract}
We report a comparative study on transmon qubit control using (i) conventional attenuated coaxial microwave line and (ii) an optical control system using modulated laser light delivered over telecommunications optical fiber to a photodiode located at the 1K stage of a dilution cryostat. During each experiment, we performed repeated measurements of the energy relaxation and coherence times of a transmon qubit using one of the control signal delivery methods. Each measurement run spanned 20 hours of measurement time and from these datasets we observe no measurable effect on coherence of the qubit compared to random coherence fluctuations. Our results open up the possibility of large scale integration of the optical qubit control system.
\end{abstract}

\maketitle

\section{Introduction}
The challenge of controlling and measuring sufficiently large qubit numbers in quantum computing has motivated various approaches, including highly dense coaxial arrays~\cite{Raicu-epj-2025}, flexible printed circuits \cite{Monarkha-APL-2024}, and signal generation using active electrical devices in the cryogenic environment \cite{underwood-PRXQuantum-2024}. Among these approaches, optical input-output through fiber coupling offers vanishingly small passive heat load from the interconnect, while potentially providing vast bandwidth for multiplexing in the optical domain~\cite{Lecocq-nature-2021}. Cryogenic fiber-optic access has enabled a variety of cryogenic optical experiments, including but not limited to cavity-enhanced quantum emitters~\cite{evans2018photon, merkel2020coherent}, quantum optomechanics~\cite{chan2011laser} and microwave-to-optics conversion~\cite{forsch2020microwave, hease2020bidirectional, mirhosseini2020superconducting, jiang2020efficient}. Recent efforts have focused on the realization of optical interfaces with superconducting qubits~\cite{mirhosseini2020superconducting, Lecocq-nature-2021, delaney2022superconducting, van2025optical, arnold2025all, warner2025coherent}. One facet of this approach focuses on optically generating qubit control pulses using a combination of room temperature (RT) electro-optic modulation of laser light followed by demodulation at cryogenic temperatures using high-speed photodiodes~\cite{Lecocq-nature-2021, xu2025manipulations, nakamura2025cryogenic}. One challenge of this approach is the thermal tradeoff arising from the conversion process between optical and microwave signals, which leads to heat generation ~\cite{Lecocq-nature-2021}. A second challenge is the question of high-frequency radiation coming from the optically coupled drive line. Standard approaches to microwave hygiene are typically incompatible with optical access, since photons with energy greater than the pair-breaking energy in a superconducting film will induce energy loss in the device. However, previous work has demonstrated that it is possible to adequately block out of band noise even when microwave signals are generated at the same thermal stage as the qubit \cite{Lecocq-nature-2021}.

In this article, we describe a practical implementation of the input control and readout-in signals for a transmon qubit,  employing optical fibers with a photodiode array inside a dilution refrigerator. We contrast these results to those obtained using all-microwave input, within the same cooldown. The article is organized as follows. First, we display the test setup with optics and a transmon qubit in a dilution refrigerator. Next, we report the time series and statistical analysis of the qubit lifetime and coherence for each configuration. We discuss the comparison between optical qubit control and all-microwave control in terms of coherence. From previous measurements that have been implemented with a blackbody noise reference, we can also estimate the effective temperature of the drive line originating from the photodiode. Finally, we comment on the scaling potential for this system from a thermal budget perspective, and assess technological readiness for standard single-qubit gate characterization based on metrics such as time stability, power efficiency, and noise. 

\section{Experimental setup}

\begin{figure}
\includegraphics[width=8cm]{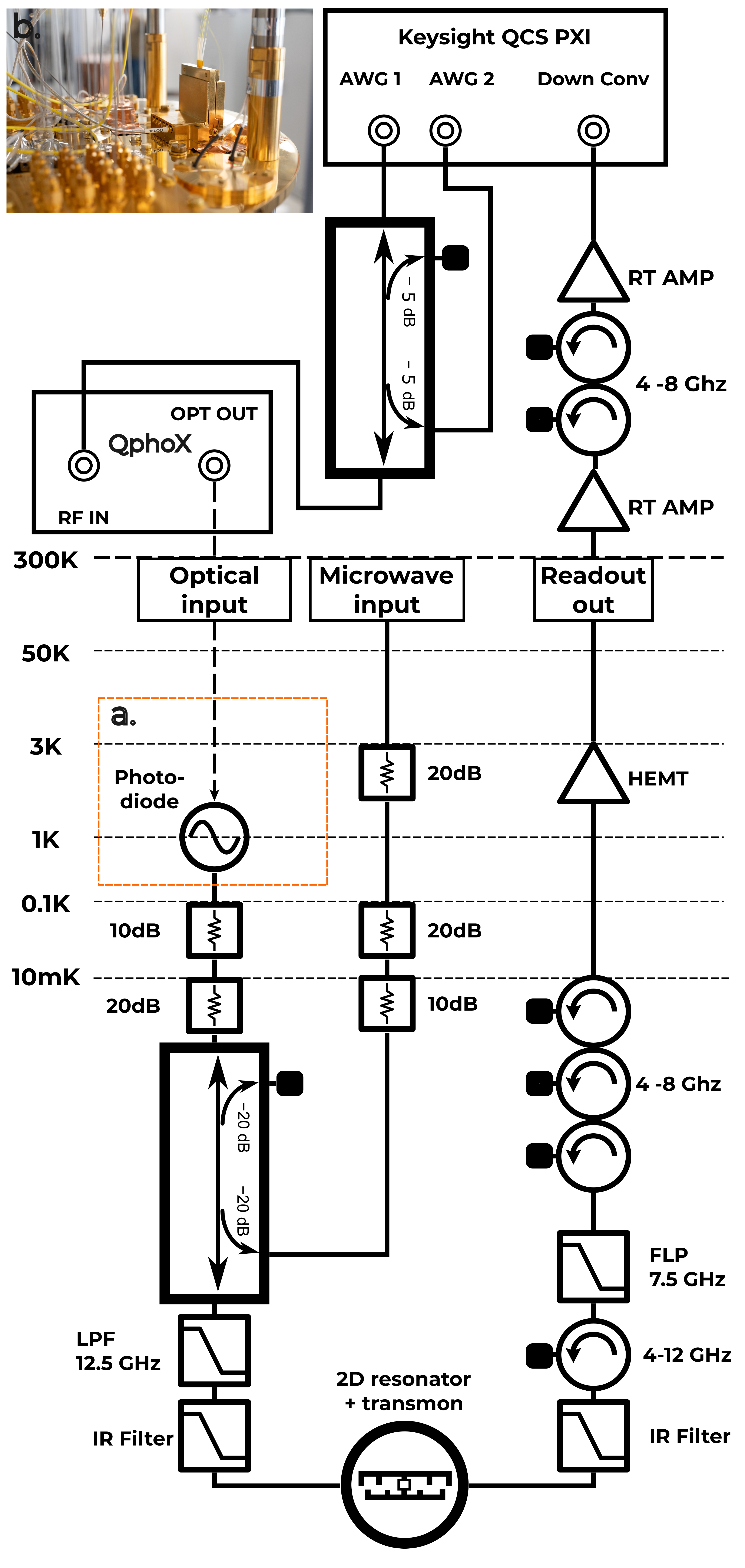}
\caption{Schematic diagram showing the measurement layout. The drive and readout signals are either delivered to the photodiode (a) over a single mode optical fiber or injected to the sample trough a coaxial attenuated input line connected to a coupled port of a directional coupler. A photographic image of the photodiode assembly is shown on the inset (b)}\label{fig:diagram} 
\end{figure}

Fig.~\ref{fig:diagram} illustrates the measurements setup used to evaluate the differences in qubit lifetime and coherence using a modulated readout/control optical signal versus a traditional all-microwave drive scheme.   The figure displays the experimental setup in a Bluefors LD400 dilution refrigerator with the qubit at a bath temperature of below \qty{10}{\milli\kelvin}, and including optical input, microwave input, and readout output.  
The optical interconnect between the room temperature control electronics and a transmon qubit is realized using the QphoX Optical Control System (OCS). Microwave frequency qubit control pulses are modulated onto a \SI{1510}{nm} laser light using an on-board laser and electro-optic intensity modulator biased at quadrature. The modulated light is transmitted from one of six optical output ports of the room temperature hardware towards the Still plate of the dilution refrigerator over single-mode telecommunications optical fiber. Via cryo-compatible fiber-optic packaging, the light is guided to a high-speed photodiode array where it is demodulated to reproduce the microwave pulses and to drive the qubit.  10 dB and 20 dB attenuators, thermalized to \qty{0.1}{\kelvin} and \qty{0.01}{\kelvin} respectively, were installed to improve thermalization of the microwave channel, in addition to a low pass filter (LPF) and an infrared filter (IRF) installed at the base temperature stage \cite{simbierowicz-prxq-2024}. Additionally, the more traditional microwave input line was connected to the qubit input using a directional coupler, enabling side-by-side comaprison of transmon dynamics for optical versus all-microwave drive. We note that when optical control was in use, both the readout and gate pulses were applied using the optical to microwave conversion. The microwave input serves only as an experimental comparison. Finally, a typical readout chain was constructed, without a parametric amplifier to avoid adding additional decoherence sources such as parametric amplifier back action noise. 

The transmon qubit (\texttt{QiB0} fabricated by ConScience, Sweden) has a common input port for both the readout stimulus and the qubit drive pulses using the so-called hanger readout resonator architecture. The device is a fixed-frequency grounded X-mon transmon that is dispersively coupled to a two-dimensional superconducting resonator, with qubit and resonator frequencies $\omega_{\textrm{q}}/\hbar = 4.5$ \unit{\giga\hertz} and $\omega_{\textrm{r}}/\hbar = 6$ \unit{\giga\hertz}, respectively. The qubit was designed with a Purcell decay time of 19.5 milliseconds. The energy participation ratio is approximately $E_\textrm{j}/E_\textrm{c} \approx 47$. The magnetic flux fluctuation contribution for such a device is considered to be negligible, and charge noise contribution is expected to be primarily to energy relaxation rather than pure dephasing via AC Stark effect induced shifts of the qubit transition frequency due to fluctuations in the thermal photon number of the readout resonator. 

\begin{figure*}
\includegraphics[width=18cm]{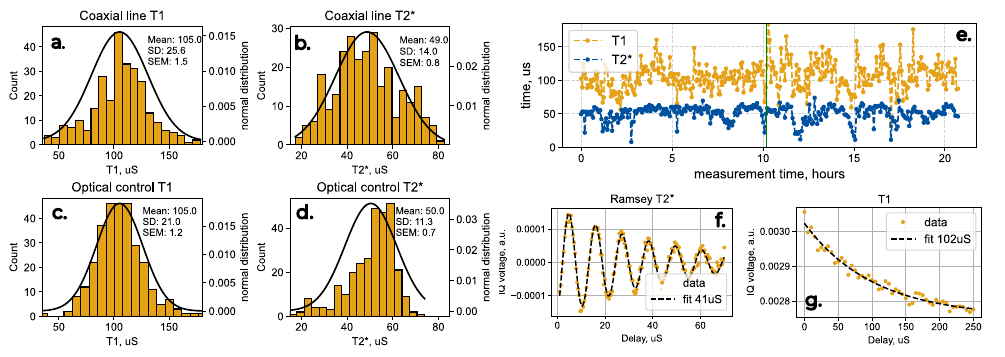}
\caption{Histograms of energy relaxation time T$_{1}$ and Ramsey T$^{*}_{2}$ accumulated over the same time period of 20 hours for control and readout input signals delivered over a microwave coaxial transmission line (a,b) or optical fiber (c,d), showing mean values, standard deviation and standard error in mean. Black line shows a fit of the normal distribution to the histogram. Time evolution of T$_{1}$,T$^{*}_{2}$ for a long interleaved measurement sequence with optical drive.}\label{fig:hist} 
\end{figure*}

\section{Measurement Results}

The main results of this study are compiled in Fig.~\ref{fig:hist} where repeated measurements were performed on the qubit, using standard interleaved  $T_{1}$ and Ramsey $T^{*}_{2}$ sequences \cite{krantz-APR-2019}. The time between measurement cycles was approximately 4 minutes First, all pulses for state preparation and readout were generated via the optical protocol, and the time series appears in Fig.~\ref{fig:hist}(e). Second, the all-microwave control and readout pulses were used to investigate potential differences and to observe possible backaction on the qubit. We compiled histograms of $T_{1}$ and $T^{*}_{2}$ from the time series of both experiments and plot the results in Fig.~\ref{fig:hist}(a,b,c,d). The time series exceeding a 20 hour window that contained heavily suppressed periods of $T_{1}$ and $T^{*}_{2}$ were not included in the analysis as we attribute those to interaction of the qubit with the slowly evolving TLS environment. The extracted mean values of pure dephasing time $T_{\phi}$ = $\frac{2 T_2 T_1}{2T_1 - T2}$ are \qty{65}{\micro\second} for optical control and \qty{64}{\micro\second} for conventional microwave control. The difference seems to be within natural statistical fluctuations of the parameter and allows to assume equivalent performance between microwave and optical control. 

\section{Discussion}

\subsection{Source stability}

In applications of quantum device characterization, stable operation over hours or days is a desirable property. By examining the peak-to-peak amplitude and horizontal asymmetry of Ramsey measurement curves, we evaluate the stability of the signal source driving the rotations. Any drift in laser power produces an under- or over-rotation of the $X_{\pi/2}$ pulses in the Ramsey sequence. For full $\ket{0}\rightarrow\ket{1}$ contrast, two ideal $X_{\pi/2}$ pulses are required, and the fringe contrast depends on whether the product of detuning and free-evolution time is an even or odd multiple of $\pi$. With over- or under-rotated pulses (e.g., $X_{\pi/4}$ or $X_{3\pi/4}$), the system still exhibits full destructive interference and returns to $\ket{0}$, but it no longer reaches a pure $\ket{1}$ state. As a result, the minima of the fringes remain at $\ket{0}$, while the maxima decrease, reducing the overall visibility and adding horizontal asymmetry to the exponential envelope of the curve. The former can be approximately quantified as a tilt coefficient B as seen in the function used to fit the Ramsey measurement curves:
\begin{gather*}
f\left( t \right) = Ae^{-\frac{t}{T_{2}}}cos\left( 2\pi f + \phi\right)+Bt+C
\end{gather*}

\begin{figure}
\includegraphics[width=8cm]{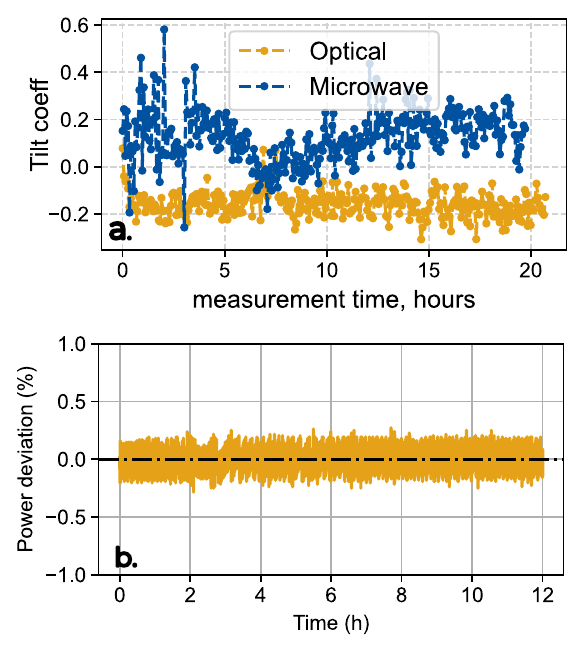}
\caption{Ramsey measurement curve tilt (a) for the time series of  $T^{*}_{2}$  measurement done with optical control or microwave control. Optical power setpoint deviation over time period of 12 hours (b).}\label{fig:stability} 
\end{figure}

By analyzing the time series of the horizontal asymmetry Fig.~\ref{fig:stability}(a) for both microwave and optical sources, we observe that the laser power is stable compared to conventional microwave driving method for all looped measurement durations of more than 20 hours. This conclusion is consistent with previous room temperature measurements of laser power stability from the same room-temperature hardware, over 12 hours seen in Fig.~\ref{fig:stability}(b). 

\subsection{Photodiode temperature and dephasing}

Next, we attempt to define an equivalent noise temperature of the photodiode, that would account for the observed pure dephasing rates $T_{\phi}$, obtained from the data in Fig.~\ref{fig:hist}(e). In an earlier cooldown with the same qubit sample and radiation shielding, as well as microwave wiring configuration, excluding the photodiode, we observed mean values of $T_{1}$ and $T^{*}_{2}$ to be \qty{120}{\micro\second} and \qty{80}{\micro\second} correspondingly that giving rise to mean pure dephasing times $T_{\phi}$ of \qty{120}{\micro\second}. This mean value differs by almost a factor of two compared to the mean values for both optical and microwave drive measured during the cooldown that includes the photodiode (both at around \qty{65}{\micro\second}). One possible explanation for such a reduction could be changes to the qubit parameters due to thermal cycling and exposure to room atmosphere between cooldowns. For example, the changed qubit transition frequency could have ended up close to a resonance frequency of a strongly coupled two-level system.

However, we can try to speculate that the reduction is possibly due to excessive thermal radiation reaching the sample through the input transmission line. We assume that for this qubit sample, the resonator photon number fluctuation is the dominant decoherence channel. In an earlier cooldown we measured the pure dephasing time  $T_{\phi}$ dependence as a function of noise temperature $T_{in}$ injected to the feedline of the qubit chip using a variable temperature noise source. To achieve that, in the setup of Fig.~\ref{fig:diagram}, the photodiode and the attenuated transmission line leading to the directional coupler were replaced with a \qty{50}{\ohm} matched resistor thermally decoupled from the Mixing Chamber (MXC) stage with precise control over its temperature. In this previous work, we used a similar thermal noise source \cite{Simbierowicz-rsi-2021} installed at the cold plate (CP) to characterize noise of a quantum limited amplifier.

From the $T_{\phi}(T_{in})$ dependence, we estimate the noise temperature $T_{in}$, reaching the qubit chip feedline, to be around \qty{100}{\milli\kelvin}, including the thermal noise contribution of the coaxial line entering the coupled port of the directional coupler calculated to be \qty{33}{\milli\kelvin}. Corrected for the line attenuation, cable and filter losses following the calculation method described in \cite{Krinner-epj-2019} (with 10dB attenuation at the CP and \qty{20}{\decibel} attenuation at the MXC, as well as IRF and LPF), this results in an estimated value of noise temperature of around \qty{24}{\kelvin}, originating from the photodiode during the measurement. In an earlier cooldown (with \qty{10}{\decibel} attenuation at CP and  \qty{10}{\decibel} at MXC), with one of the RF ports of the photodiode connected to an unattenuated RT return line, the extracted noise power at qubit chip input is estimated to be around ~\qty{500}{\milli\kelvin} which corresponds to about \qty{50}{\kelvin} at Still stage. A higher value could be due to the unattenuated return line present in that cooldown. 

Notably, the values of estimated noise temperature for both cooldowns were independent of whether the laser signal  was on or off. This suggests that if the interaction of the qubit with its environment has not changed significantly between the cooldowns, then the nature of the measured elevated noise temperature is not related to elevated photodiode chip temperature as a result of laser signal power dissipation or its shot noise and should be investigated further. One possible explanation of the elevated noise temperature could be  room temperature thermal radiation leaking through the pass bands of the optical fiber and heating up the chip.

As we speculate about further scalability of the solution we see a possibility to optimize cooling power of the dilution refrigerator systems for specific signal generation stages to account for desired channel density. Additionally the optical power duty cycle can be adjusted to minimize active heat load, that is the laser power can be kept on exclusively when a microwave pulse is being sent. For instance the \qty{50}{\micro\watt} optical power value we used in this manuscript will be reduced to \qty{5}{\micro\watt} at a duty cycle of 10\%. 

\subsection{Thermal budget and scaling}

When looking at active thermal load produced by the photodiode we expect it to be the same as delivered optical power over a cycle. The Still stage of a Bluefors LD400 is usually heated with \qty{7}{\milli\watt} using an electrical heater to achieve optimal helium isotope mixture flow through the cryostat. We did not notice a measurable increase in the Still temperature when turning the optical power of ~\qty{50}{\micro\watt} on or off with 100\% duty cycle. 
To better illustrate  the scalability of the optical approach versus the traditional coaxial-based scenario we performed the following thermal simulations. We modeled a Bluefors XLD1000s dilution refrigerator, the standard choice for large-scale quantum computing architectures, and computed the thermal loads and steady state flange temperatures for the three different control and measurement configurations: \textbf{(1)} all-coax wiring, \textbf{(2)} optical + coax wiring,  \textbf{(3)} \ optical + superconducting coax wiring. In all of the three scenarios we assume 840 control lines and 168 readout lines; these sum up to 1008 lines, the maximum amount of coaxial lines in high density configuration that can be installed in an XLD1000s. The readout lines are the same in all simulations: 0.86 SCuNi coaxial wires connecting from room temperature feedthrough to the 4K flange and 0.86 NbTi coaxial wires from the 4K to the MXC flange. The control lines are 0.86 SCuNi coaxes from RT to MXC in \textbf{(1)}, optical fibers from RT to the Still flange and 0.86 SCuNi coaxes from the Still to the MXC flange in \textbf{(2)}, and finally optical fibers from RT to the Still flange and 0.86 NbTi coaxes from the Still to the MXC flange. In the simulation we apply RF power to all control lines at a 33 \% duty cycle and the input RF power at the RT connection is chosen as to result in the same power delivered at MCX, about \num{-63} dBm, in all the 3 scenarios. For the all-coax simulation \textbf{(1)} we change the attenuation scheme slightly from the experiment and set \qty{20}{\decibel} of attenuation at \qty{4}{\kelvin}, CP, and MXC flange. Finally, we assume that the refrigerator is run at a still heating power of about \qty{80}{\milli\watt}, resulting in a high molar flow regime, about 2200 \textmu mol/s, which may be necessary to improve the MXC cooling power when operating heavily loaded systems.

\begin{table}
    \centering
        \caption{Flange thermal loads for the three different wiring configurations assuming 840 control and 168 readout lines. (N) refers to normal coaxes and (S) to superconducting coaxes. The Still thermal load does not include the Still heater contribution that must be added to keep the molar flow stable and the Still temperature constant.}
    \begin{tabular}{|c|c|c|c|}
    \hline
      Load  & All-coax (N) & Optical-coax (N) & Optical-coax (S) \\
        \hline
       50K (W)  & 10.188 & 4.985 & 4.985\\
         \hline
       4K (W)  &  0.897 & 0.285 & 0.285 \\
         \hline
       Still (\si{\milli\watt}) & 2.853 & 13.949 &  13.939 \\
       \hline
       CP (\si{\micro\watt}) & 2502.807 & 777.158 &  315.822 \\
       \hline
       MXC (\si{\micro\watt}) &  48.607 & 33.190 &  32.060 \\
         \hline
    \end{tabular}

    \label{tab:thermal_loads}
\end{table}

\begin{table}
    \centering
        \caption{Flange temperature estimates for a Bluefors XLD1000s with 2 pulse tubes PT420, for the three different wiring configuration. (N) refers to normal coaxes and (S) to superconducting coaxes. The Still temperature is assumed fixed by supplying the flange with the extra power from the heater needed to keep the molar flow stable.}
    \begin{tabular}{|c|c|c|c|}
    \hline
      Temperature  & All-coax (N) & Optical-coax (N) & Optical-coax (S)  \\
        \hline
       50K (K)  & 36.038 & 35.349 & 35.349\\
         \hline
       4K (K)  & 3.590 & 2.998 & 2.988 \\
         \hline
       Still (K) & 1.4 & 1.4 &  1.4 \\
       \hline
       CP (\si{\milli\kelvin}) & 245.354 & 155.981 &  121.412 \\
       \hline
       MXC (\si{\milli\kelvin}) &  22.735 & 19.409 &  19.143 \\
         \hline
    \end{tabular}
    \label{tab:temps}
\end{table}

The thermal loads are reported in Table~\ref{tab:thermal_loads}. The thermal load at Still flange does not include the Still heater contribution, that needs to be added in order to reach a Still temperature of about \qty{1.4}{\kelvin}. The increase in this flange's thermal load when switching to optical control is clear compared to the all-coax case and it is dominated by the optical power dissipated by the photodiode which, in our simulation amounts to about 14 mW, assuming 840 photodiodes and a 33\% duty cycle. This is higher than the Still heater power needed to run an XLD1000s at lower flow, which is approximately 12 mW. Thus, compared to the traditional all-coax approach, the optical approach may push dilution refrigerator (DR) operations towards a higher flow regime, resulting in higher cooling power at lower stages. The lowest thermal loads are for configuration \textbf{(3)}, where no conductive heat loads are present at the upper stages and reduction of the loads at CP and MXC can also be observed. The load at CP for configuration \textbf{(2)} is higher than for configuration 3 due to the conductor losses of SCuNi. At MXC this difference nearly disappears as the \qty{20}{\decibel} attenuators dominate the heating mechanism. The resulting flange temperature for an XLD1000s system equipped with two pulse tubes PT 420 estimates are displayed in Table~\ref{tab:temps} and they also reflect the better thermal performance of the optical control over the traditional coaxial approach. We conclude this section by noting if we modify the scenario \textbf{(3)} (the optical + superconducting coax) by installing the photodiodes at the 4K flange instead of the Still flange and running the optical fibers from RT to 4K and superconducting wires from 4K to MXC, the heat load on the Still reduces dramatically while the load on the 4K flange increases marginally. The resulting 4K temperature would sit at around 3.012 K, while the CP and MXC flanges would be at 107.272 mK and 18.663 mK, respectively.   

\section{Conclusion}

In summary, we have performed a comparative study on hybrid optical-microwave and all-microwave based qubit control techniques. Our results show (i) no statistically significant difference in qubit coherence times between the two techniques and (ii) a potential advantage in terms of thermal management, depending on the control wiring configuration, for the optical case when simulating a hypothetical XLD1000s refrigerator equipped with the maximum possible number of control and readout lines.
In the future, a system may be built based on a hybrid optical-microwave control and measurement infrastracture. In such a system the control signals are delivered to the \qty{1}{\kelvin} or \qty{4}{\kelvin} stage of the cryostat via optical fibers, converted to microwave domain with the photodiode arrays and delivered to the QPU through attenuated superconducting coaxial or flexible microwave transmission lines. The cooling power of the \qty{1}{\kelvin} or \qty{4}{\kelvin} stages can be improved to account for higher channel density by introducing more powerful cryocooler solutions and adjustments to DR unit design.  

\section*{Acknowledgments}
The Bluefors quantum testbed is supported by the EU Flagship on Quantum Technology project OpenSuperQPlus100 (HORIZON-CL4-2022-QUANTUM-01-SGA, 101113946). QphoX acknowledges Cristobal Ferrer and Katie Helsby for their contribution to the development of the optical control system.

\section*{Competing interests}

R.H.K., M.K., E.C., B.d.Z., M.Z., K.P., A.P., F.D.M., B.S., S.G., T.C.v.T. and R.S. are or have been employed by QphoX B.V. and are, have been or may in the future be participants in incentive stock plans at QphoX B.V.

\bibliography{bibliography}

\end{document}